\documentclass[runningheads,a4paper]{llncs}

\usepackage{amsmath,amssymb,amsbsy,latexsym}
\usepackage{graphicx,epic,eepic}
\usepackage{cite}
\usepackage{epstopdf}
\usepackage{longtable}
\usepackage{url}

\begin{document}

\mainmatter  

\title{Cooperative Game Theory Approaches\\ for Network Partitioning}

\titlerunning{Network Partitioning as Cooperative Game}

%
%

\author{Konstantin E. Avrachenkov\inst{1}, Aleksei Yu. Kondratev\inst{2}, Vladimir V. Mazalov\inst{2}}
\authorrunning{K.E.Avrachenkov,  A.Y.Kondratev and  V.V.Mazalov}

\institute{INRIA, 2004 Route des Lucioles, Sophia-Antipolis, France
\email{k.avrachenkov@sophia.inria.fr}
\and
Institute of Applied Mathematical Research, Karelian Research Center,
\\Russian Academy of Sciences, 11, Pushkinskaya st., Petrozavodsk, Russia, 185910\\
\email{vmazalov@krc.karelia.ru}}

\maketitle

%
%

\begin{abstract}
The paper is devoted to game-theoretic methods for community detection in networks.
The traditional methods for detecting community structure are based on selecting
denser subgraphs inside the network. Here we propose to use the methods of cooperative
game theory that highlight not only the link density but also the mechanisms of
cluster formation. Specifically, we suggest two approaches from cooperative game
theory: the first approach is based on the Myerson value, whereas the second
approach is based on hedonic games. Both approaches allow to detect clusters with
various resolution. However, the tuning of the resolution parameter in the hedonic
games approach is particularly intuitive. Furthermore, the modularity based approach
and its generalizations can be viewed as particular cases of the hedonic games.
\keywords{Network partitioning, community detection, cooperative games,
Myerson value, hedonic games.}
\end{abstract}

\section{Introduction}

Community detection in networks or network partitioning is a very important topic
which attracted the effort of many researchers. Let us just mention several main
classes of methods for network partitioning. The first very large class is based
on spectral elements of the network matrices such as adjacency matrix and Laplacian
(see e.g., the survey \cite{L07} and references therein). The second class of methods,
which is somehow related to the first class, is based on the use of random walks
(see e.g., \cite{Aetal08,ACN14,D00,MS01,newman,PL06} for the most representative works in this research direction.)
The third class of approaches to network partitioning is based on methods from
statistical physics \cite{BWD96,RAK07,RB06}. The fourth class, which is probably most related to
our approach, is based on the concept of modularity and its various generalizations
\cite{Betal08,girvan,newman1,Wetal10}.
For a very thorough overview of the community detection methods we recommend
the survey \cite{F10}.

In essence, all the above methods (may be with some exception of the statistical
physics methods), try to detect denser subgraphs inside the network and do not
address the question: what are the natural forces and dynamics behind the formation of network
clusters. We feel that the game theory, and in particular, cooperative game
theory is the right tool to explain the formation of network clusters.

In the present work, we explore two cooperative game theory approaches to
explain possible mechanisms behind cluster formation. Our first approach is
based on the Myerson value in cooperative game theory, which particularly
emphasizes the value allocation in the context of games with interactions between
players constrained by a network. The advantage of the Myerson value is in taking
into account the impact of all coalitions. We use the efficient method developed
in \cite{maz_tr} and \cite{fundam} based on characteristic functions to calculate
quickly the Myerson value in the network. We would like to mention that in \cite{fundam}
a network centrality measure based on the Myerson value was proposed. It might be interesting
to combine node ranking and clustering based on the same approach such as the Myerson
value to analyze the network structure.

The second approach is based on hedonic games,
which are games explaining the mechanism behind the formation of coalitions.
Both our approaches allow to detect clusters with varying resolution and thus
avoiding the problem of the resolution limit \cite{FB07,Letal09}. The hedonic game
approach is especially well suited to adjust the level of resolution as the
limiting cases are given by the grand coalition and maximal clique decomposition,
two very natural extreme cases of network partitioning. Furthermore, the
modularity based approaches can be cast in the setting of hedonic games.
We find that this gives one more, very interesting, interpretation of the
modularity based methods.

Some hierarchical network partitioning methods based on tree hierarchy,
such as \cite{girvan}, cannot produce a clustering on one resolution level
with the number of clusters different from the predefined tree shape.
Furthermore, the majority of clustering methods require the number of clusters
as an input parameter.
In contrast, in our approaches we specify the value of the resolution parameter
and the method gives a natural number of clusters corresponding to the given
resolution parameter.

In addition, our approach easily works with multi-graphs, where several edges (links)
are possible between two nodes. A multi-edge has several natural interpretations
in the context of social networks. A multi-edge can represent: a number of telephone
calls; a number of exchanged messages; a number of common friends; or a number
of co-occurrences in some social event.

The paper is structured as follows: in the following section we formally define
network partitioning as a cooperative game. Then, in Section~3 we present our
first approach based on the Myerson value. The second approach based on the
hedonic games is presented in Section~4. In both Sections~3~and~4 we provide
illustrative examples which explain the essence of the methods. Finally, Section~5
concludes the paper with directions for future research.


\section{Network partitioning as a cooperative game}

Let $g=(N,E)$ denote an undirected multi-graph consisting of the set of nodes \textit{N} and the set of edges \textit{E}. We denote a link between node $i$ and node $j$ as $ij$.
The interpretation is that if $ij\in E$, then the nodes $i\in N$ and $j\in N$ have a connection in network $g$,
while $ij\notin E$, then nodes $i$ and $j$ are not directly connected. Since we generally consider a multi-graph,
there could be several edges between a pair of nodes. Multiple edges can be interpreted for instance as a number of telephone
calls or as a number of message exchanges in the context of social networks.

We view the nodes of the network as players in a cooperative game.
Let \linebreak $N(g)=\{i:\exists j \mbox{ such that } ij\in g\}$.
For a graph $g$, a sequence of different nodes $\{i_1,i_2,\dots ,i_k\},\ k\ge 2$, is a path connecting $i_1$ and $i_k$ if for all $h=1,\dots ,k-1$, $i_hi_{h+1}\in g$. The length of the path \textit{l} is the number of links in the path, i.e. $l=k-1$. The length of the shortest path connecting $i$ and $j$ is distance between $i$ and $j$. Graph $g$ on the set \textit{N} is connected graph if for any two nodes $i$ and $j$ there exists a path in $g$ connecting $i$ and $j$.

We refer to a subset of nodes $S \subset N$ as a coalition. The coalition \textit{S} is connected if any two nodes
in \textit{S} are connected by a path which consists of nodes from $S$.
The graph $g'$ is a component of $g$, if for all $i\in N(g')$ and $j\in N(g')$, there exists a path in $g'$ connecting
$i$ and $j$, and for any $i\in N(g')$ and $j\in N(g)$, $ij\in g$ implies that $ij\in g'$. Let $N|g$ is the set of all components in $g$ and let $g|S$ is the subgraph with the nodes in $S$.

Let $g-ij$ denote the graph obtained by deleting link \textit{ij} from the graph  $g$ and $g+ij$
denote the graph obtained by adding link \textit{ij} to the graph $g$.

The result of community detection is a partition of the network $(N,E)$ into
subsets (coalitions) $\{S_1,...,S_K\}$ such that $S_k \cap S_l=\emptyset, \forall k,l$ and
$S_1\cup...\cup S_K=N$.  This partition is {\it internally stable} or {\it Nash stable} if for any player from
coalition $S_k$ it is not profitable to join another (possibly empty) coalition $S_l$.
We also say that the partition is {\it externally stable} if for any player $i\in S_l$ for whom it is benefitial
to join a coalition $S_k$ there exists a player $j\in S_k$ for whom  it is not profitable
to include there player $i$. The payoff definition and distribution will be discussed in the following
two sections.

\section{Myerson cooperative game approach}

In general, cooperative game of \textit{n} players is a pair $<N,v>$ where $N=\{1,2,\dots ,n\}$ is
the set of players and \textit{v}: $2^N\rightarrow R$ is a map prescribing for a coalition $S\in 2^N$
some value $v(S)$ such that $\textit{v}(\emptyset) = 0$. This function $v(S)$ is the total utility
that members of $S$ can jointly attain.
Such a function is called the characteristic function of cooperative game \cite{maz_gt}.

Characteristic function (payoff of coalition $S$) can be determined in different ways.
Here we use the approach of \cite{jac2005,jac,fundam,maz_tr}, which is based on discounting directed paths.
The payoff to an individual player is called an imputation. The imputation in this cooperative game
will be Myerson value \cite{my,fundam,maz_tr}.

Let $<N, v>$ be a cooperative game with partial cooperation presented by graph $g$ and characteristic function $v$.
An allocation rule $Y$ describes how the value associated with the network is distributed to the individual players.
Denote by $Y_i(v,g)$ the value allocated to player $i$ from graph $g$ under the characteristic function $v$.

Myerson proposed in \cite{my} the allocation rule
$$
Y(v,g)=(Y_1(v,g), \dots, Y_n(v,g)),
$$
which is uniquely determined by the following two axioms:

\textbf{A1}. If \textit{S} is a component of $g$ then the members of the coalition \textit{S} ought to allocate
to themselves the total value $v(S)$ available to them, i.e $\forall S\in N|g$
\begin{equation}
 \sum_{i\in S} Y_i(v, g)=v\left(S\right).
 \label{eq1}
\end{equation}

\textbf{A2}. $\forall g,\ \forall ij\in g$ both players $i$ and $j$ obtain equal payoffs after adding or deleting a link $ij$,
\begin{equation}
 Y_i\left(v,g\right)-Y_i\left(v,g-ij\right)=Y_j\left(v,g\right)-Y_j\left(v,g-ij\right).
 \label{eq2}
\end{equation}
Let us determine the characteristic function by the following way
\[v_g\left(S\right)=\sum_{K\in S|g} v\left(K\right).\]
Then the Myerson value can be calculated by the formula
\begin{equation}
Y_i\left(v,\ g\right)=\sum_{S\subset N\backslash \{i\}}{(v_g\left({S\cup i}\right)-v_g\left(S\right))}\,\frac{s!\left(n-s-1\right)!}{n!},
\label{eq3}
\end{equation}
where $s=\left|S\right|$ and $n=\left|N\right|.$

Let us determine the characteristic function which is  determined by the
scheme proposed by Jackson \cite{jac}: every direct connection gives to coalition \textit{S} the impact \textit{r},
where $0\leq r \leq 1$. Moreover, players obtain an impact from indirect connections. Each path of length 2 gives to coalition \textit{S} the impact $r^2$,
a path of length  3 gives to coalition the impact $r^3$, etc. So, for any coalition  \textit{S} we obtain
\begin{equation}
 v\left(S\right)=a_1r+a_2r^2+\dots +a_kr^k+\dots +a_Lr^L=\sum^L_{k=1}{a_k}r^k,
 \label{eq4}
\end{equation}
where \textit{L} is a maximal distance between two nodes in the coalition; $a_k$ is the number of paths
of length \textit{k} in this coalition. Set
\[v(i)=0,\ \forall i\in N.\]

In \cite{maz_tr} it was proven that the Myerson value can be found by
the following simple procedure of allocation the general gain $v(N)$ to each player $i\in N$:

\textbf{Stage 1.} Two direct connected players together obtain \textit{r}. Individually, they would receive nothing.
So, each of them receives at least $r/2$. If player \textit{i} has some direct connections then
she receives the value $r/2$ times the number of paths of length 1 which contain the node \textit{i}.

\textbf{Stage 2.} Three connected players obtain $r^2$, so each of them must receive $r^2/3$, and so on.

\noindent Arguing this way, we obtain the allocation rule of the following form:
\begin{equation}
Y_i\left(v,g\right)=\frac{a^i_1}{2}r+\frac{a^i_2}{3}r^2+\dots +\frac{a^i_L}{L+1}r^L=\sum^L_{k=1}{\frac{a^i_k}{k+1}}r^k,
\label{eq5}
\end{equation}
where $a^i_k$ is the number of all paths of length \textit{k} which contain the node \textit{i}.

\medskip

\noindent
{\bf Example 1.} Consider network of six nodes presented in Fig. 1. Below we show how to calculate
characteristic function for different coalitions.

\begin{figure}
\centering
\includegraphics[height=3cm]{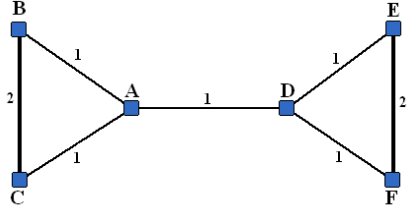}
\caption{Network of six nodes.}
\label{fig:example1}
\end{figure}

\noindent
For the network $N= \{A,B,C,D,E,F\}$ we find \textit{L} = 3, $a_1=9$, $a_2=4$, $a_3=4$.
Consequently, the value  of grand-coalition is
$$
v\left(N\right)=9r+4r^2+4r^3.
$$
For coalition $S=\left\{A,B,C,D\right\}$ we have \textit{L} = 2, $a_1=5$, $a_2=2$ and we obtain
$$
v\left(S\right)=5r+2r^2.
$$
This way we can calculate the values of characteristic function for all coalitions $S\subset N$.
After that we can find the Myerson vector.

\medskip

\noindent
{\bf Example 1 (ctnd).}
Let us calculate the Myerson value for player A  in Example~1 using the allocation rule (\ref{eq5}).
Mark all paths which contain node A.
The paths of length 1 are: \{A,B\}, \{A,C\}, \{A,D\}, hence $a^A_1=3$.
The paths of length 2 are: \{B,A,C\}, \{B,A,D\}, \{C,A,D\}, \{A,D,E\}, \{A,D,F\}, so $a^A_2=5$.
The paths of length 3: \{B,A,D,E\}, \{B,A,D,F\},  \{C,A,D,E\}, \{C,A,D,F\}, so  $a^A_3=4$.
 Consequently,
\[ Y_A=\frac{3}{2}r+\frac{5}{3}r^2+r^3.\]

\medskip

Thus, we can propose the following algorithm for network partitioning based on the Myerson value:
Start with any partition of the network $N=\{S_1,\ldots,S_K\}$. Consider a coalition
$S_l$ and a player $i\in S_l$. In cooperative game with partial cooperation presented by the graph
$g|S_l$ we find the Myerson value for player $i$,  $Y_i(g|S_l)$. That is reward of player $i$
in coalition $S_l$.  Suppose that player $i$ decides to join the coalition $S_k$. In the new
cooperative game with partial cooperation presented by the graph $g|S_k\cup i$ we find the Myerson value
$Y_i(g|S_k\cup i)$.  So, if for the player $i\in S_l:$ $Y_i(g|S_l)\ge Y_i(g|S_k\cup i)$
then player $i$ has no incentive to join to new coalition $S_k$, otherwise the player changes
the coalition.
The partition  $N=\{S_1,\ldots,S_K\}$ is the Nash stable if for any player there is no incentive
to move from her coalition. Notice that for unweighted graphs  the definition of the Myerson value
implies that for any coalition it is always beneficial to accept a new player (of course, for the player herself it might not be
profitable to join that coalition), the Nash stability (internal stability)
in this game coincides with the external stability.

\medskip

\noindent
{\bf Example 1 (ctnd)}.
Let us clarify this approach on the network
$$
N=\{A,B,C,D,E,F\}
$$
presented in Fig. 1.
Natural way of partition here is  $\{S_1=(A,B,C), S_2=(D,E,F)\}$. Let us determine under which condition
this structure will present the stable partition.

Suppose that characteristic function is determined by (\ref{eq4}).
For coalition $S_1$ the payoff  $v(S_1)=4r$.   The payoff of player A  is $Y_A(g|S_1)=r$.
Imagine that player $A$ decides to join the coalition $S_2$.

Coalition $S_2\cup A$ has payoff $v(S_2\cup A)=5r+2r^2$. The  imputation in this coalition is
$Y_A(g|S_2\cup A)=r/2+2r^2/3,  Y_D(g|S_2\cup A)=3r/2+2r^2/3, Y_E(g|S_2\cup A)=Y_F(g|S_2\cup A)=3r/2+r^2/3$.
We see that for player $A$ it is profitable to join this new
coalition if $r/2+2r^2/3>r$, or $r>3/4$. Otherwise, the coalitional structure is stable.

Thus, for the network in Fig. 1 the Myerson value approach will give the partition
$\{S_1=(A,B,C), S_2=(D,E,F)\}$ if $r < 3/4$ and, otherwise, it leads to the grand coalition.
This example already gives a feeling that the parameter $r$ can be used to tune the resolution
of network partitioning. Such tuning will be even more natural in the ensuing approach.

\section{Hedonic coalition game approach}

There is another game-theoretic approach for the partitioning of a
society into coalitions based on the ground-breaking work \cite{Bogomolnaia}.
We apply the framework of Hedonic games \cite{Bogomolnaia} to network
partitioning problem, particularly, specifying the preference function.

Assume that the set of players $N=\{1,\ldots, n\}$ is divided into $K$ coalitions:
$\Pi=\{S_1,\ldots,S_K\}$. Let $S_\Pi(i)$ denote the coalition $S_k\in\Pi$ such that $i\in S_k$. A player $i$
preferences are represented by a complete, reflexive and
transitive binary relation $\succeq_i$ over the set $\{S\subset N:
i\in S\}$.  The preferences are additively separable \cite{Bogomolnaia} if there exists a
value function $v_i:N\rightarrow \mathbb{R}$ such that $v_i(i)=0$
and
\begin{equation*}
S_1\succeq_i S_2 \Leftrightarrow \sum\limits_{j\in S_1}{v_i(j)}\geq
\sum\limits_{j\in S_2}{v_i(j)}.
\end{equation*}

The preferences  $\{v_i,  i\in N\}$ are symmetric, if
$v_i(j)=v_j(i)=v_{ij}=v_{ji}$ for all $i,j\in N$. The symmetry
property defines a very important class of Hedonic games.

As in the previous section, the network partition $\Pi$ is {\it Nash stable},
if $S_\Pi(i)\succeq_i S_k\cup \{i\}$ for all $i\in N, S_k\in \Pi\cup \{\emptyset\}$.
In the Nash-stable partition,  there is no player who wants to leave her coalition.

A potential of a coalition partition $\Pi=\{S_1,\ldots,S_K\}$ (see \cite{Bogomolnaia}) is
\begin{equation}
\label{potential}
P(\Pi)=\sum_{k=1}^{K} P(S_k)=\sum\limits_{k=1}^K{\sum\limits_{i,j\in S_k}{v_{ij}}}.
\end{equation}

Our method for detecting a stable community structure is based on
the following better response type dynamics:

\medskip

Start with any partition of the network $N=\{S_1,\ldots,S_K\}$. Choose any player $i$
and any coalition $S_k$ different from $S_\Pi(i)$. If $S_k\cup \{i\} \succeq_i S_\Pi(i)$,
assign node $i$ to the coalition $S_k$; otherwise, keep the partition unchanged and choose
another pair of node-coalition, etc.

Since the game has the potential (\ref{potential}), the above algorithm is guaranteed to converge
in a finite number of steps.

\medskip

\noindent {\bf Proposition 1.} \textit{If players' preferences are
additively separable and symmetric ($v_{ii}=0, v_{ij}=v_{ji}$ for
all $i,j\in N$), then the coalition partition $\Pi$ giving a local maximum of
the potential $P(\Pi)$ is the Nash-stable partition.}

\medskip

One natural way to define a symmetric value function
$v$ with a parameter $\alpha\in[0,1]$ is as follows:
\begin{equation}\label{value}
v_{ij}= \left\{ \begin{array}{cc} 1-\alpha, & (i,j)\in E, \\
-\alpha, & (i,j)\notin E, \\ 0, & i=j.
\end{array} \right.
\end{equation}

For any subgraph $(S,E|S)$, $S\subseteq N$, denote $n(S)$ as the number
of nodes in $S$, and $m(S)$ as the number of edges in $S$. Then, for
the value function (\ref{value}), the potential (\ref{potential})
takes the form
\begin{equation}\label{potential2}
P(\Pi)=\sum\limits_{k=1}^K{\left( m(S_k)-\frac{n(S_k)(n(S_k)-1)
\alpha}{2} \right)}.
\end{equation}

We can characterize the limiting cases $\alpha \to 0$ and $\alpha \to 1$.

\medskip

\noindent {\bf Proposition 2.} \textit{If $\alpha=0$, the grand coalition partition $\Pi_N=\{ N \}$
gives the maximum of the potential $(8)$. Whereas if $\alpha \to 1$, some local maximum
of $(8)$ corresponds to a network decomposition into disjoint maximal cliques, given such 
decomposition exists. $($A maximal clique is a clique which is not contained in another clique.$)$}

\medskip

\noindent {\bf Proof:}
It is immediate to check that for $\alpha=0$ the grand coalition partition $\Pi_N=\{ N \}$ gives the maximum of the
potential (\ref{potential2}), and  $P(\Pi_N)=m(N)$.

For values of $\alpha$ close to 1, the partition into maximal cliques  $\Pi=\{S_1,\ldots,S_K\}$
gives the maximum of  (8). Indeed, assume that a player $i$ from the  clique $S_{\Pi}(i)$
of the size $m_1$ moves to a clique $S_j$ of the size $m_2<m_1$. The player  $i\in S_{\Pi}(i)$ and $S_j$
are connected by at most $m_2$  links.   The impact on $P(\Pi)$ of this movement is not higher than
$$
m_2(1-\alpha)-(m_1-1)(1-\alpha)\le 0.
$$
Now, suppose that  player $i$ from the  clique $S_{\Pi}(i)$  moves to a clique $S_j$ of the size $m_2\ge m_1$.
 The player  $i\in S_{\Pi}(i)$ is connected with the  clique  $S_j$  by at most $m_2-1$  links.
Otherwise,  it contradicts the fact that $\Pi$ is maximal clique cover and  the clique $S_j$ can be increased by adding of $i$.
If $i$ has an incentive to move from  $S_{\Pi}(i)$ to the  clique  $S_j$, then for new partition the sum (8)
would be not higher than for partition $\Pi$ by
$$
m_2-1-m_2\alpha -(m_1-1)(1-\alpha)=m_2-m_1-\alpha(m_2-m_1+1).
$$
For $\alpha$ close to 1, this impact is negative, so there is no incentive to join the coalition $S_j$.

\medskip

The grand coalition and the maximal clique decomposition are two extreme partitions into communities.
By varying the parameter $\alpha$ we can easily tune the resolution of the community detection algorithm.

\medskip

\noindent {\bf Example 2.} Consider  graph $G=G_1\cup G_2 \cup G_3
\cup G_4$, which consists of $n=26$ nodes and $m=78$ edges (see Fig.~2.) This graph
includes 4 fully connected subgraphes: $(G_1,8,28)$   with 8 vertices connected
 by 28 links, $(G_2,5,10)$, $(G_3,6,15)$ and
$(G_4,7,21)$. Subgraph $G_1$ is connected with $G_2$ by 1 edge,
$G_2$ with $G_3$ by 2 edges, and $G_3$ with $G_4$ by 1 edge.

\begin{figure}
\centering
\includegraphics[height=5cm]{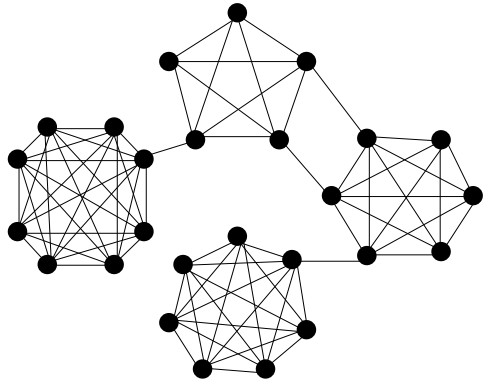}
\caption{Graph with four fully connected subgraphs.}
\label{fig:example2}
\end{figure}

Firstly, find the potentials (\ref{potential2}) for large-scale
decompositions of $G$ for any parameter $\alpha\in [0,1]$. It is
easy to check, that $P(G)=78-325\alpha$, $P(\{G_1, G_2\cup G_3\cup
G_4\})=77-181\alpha$, $P(\{G_1, G_2\cup G_3, G_4\} )=76-104\alpha$,
$P(\{ G_1, G_2, G_3, G_4\} )=74-74\alpha$.

Other coalition partitions give smaller potentials: $P(\{ G_1 \cup
G_2, G_3\cup G_4\})=76-156\alpha<76-104\alpha$, $P(\{ G_1 \cup G_2
\cup G_3, G_4\})=77-192\alpha<77-181\alpha$, $P(\{ G_1, G_2, G_3
\cup G_4 \})=75-116\alpha<76-104\alpha$, $P(\{ G_1 \cup G_2, G_3,
G_4 \})=75-114\alpha<76-104\alpha$.

We solve a sequence of linear inequalities in order to find
maximum of the potential for all $\alpha\in[0,1]$. The result is
presented in the table below.

\begin{center}
{\small\textit{Nash-stable coalition partitions in Example~2.}
\\* \vspace{3mm}
\begin{tabular}{|c|c|c|}
\hline
 $\alpha$  & coalition  partition & potential   \\
\hline
$[0,1/144]$ & $G_1 \cup G_2 \cup G_3 \cup G_4$ &  $78-325\alpha$ \\
\hline
$[1/144,1/77]$ & $G_1, G_2 \cup G_3 \cup G_4$ &  $77-181\alpha$ \\
\hline
$[1/77,1/15]$ & $G_1, G_2 \cup G_3, G_4$ & $76-104\alpha$ \\
\hline
$[1/15,1]$ & $G_1, G_2, G_3, G_4$ & $74-74\alpha$ \\
\hline
\end{tabular}
}
\end{center}

\medskip

\noindent {\bf Example 1 (ctnd).} Note that for the unweighted version of the network example presented in Fig.~1,
there are  only two stable partitions:
$\Pi=N$ for small values of $\alpha\le 1/9$ and $\Pi=\{\{A,B,C\},\{D,E,F\}\}$  for $\alpha>1/9$.

\medskip
\noindent {\bf Example 3.} Consider the popular example of the social network from Zachary karate club (see Fig.~3).
In his study \cite{Z77}, Zachary observed 34 members of a karate club over a period of two years.  Due to a disagreement developed between the administrator of the club and the club's instructor there appeared two new clubs associated with the instructor (node 1) and administrator (node 34)
of sizes 16 and 18, respectively.

\begin{figure}
\centering
\includegraphics[height=4cm]{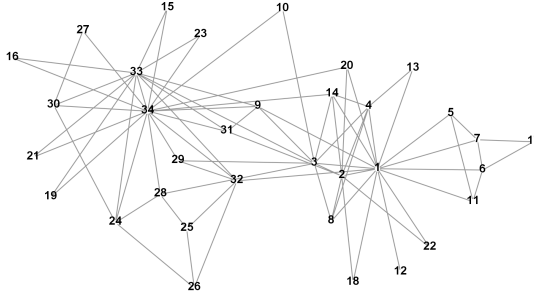}
\caption{Zachary karate club network.}
\label{fig:example3}
\end{figure}

The authors of \cite{girvan} divide the network into two groups of roughly equal size using the hierarchical
clustering tree. They show that this split corresponds almost perfectly with the actual division of the club members following the break-up. Only one node, node 3, is classified incorrectly.

Let us now apply the hedonic game approach to the karate club network. We start from the final partition
$N=\{S_{15}, S_{19}\}$, which was obtained in \cite{girvan}.
We calculate the potential for grand-coalition $ P(N)=78-561\alpha $ and
for partition $ P(S_{15},S_{19})=68-276\alpha$.  From the equation $ P(N)= P(S_{15},S_{19})$
we obtain the cutoff point $\alpha=2/57$.
So, if $\alpha <2/57$, $P(N)$ is larger than $P(S_{15},S_{19})$, so partition $\{S_{15},S_{19}\}$  is not Nash-stable.
For $\alpha=2/57$ the potential increases if the node 3 moves from $S_{19}$ to $S_{15}$.
For the new partition $P(S_{16},S_{18})=68-273\alpha$.
Comparing with potential of the grand coalition we obtain $\alpha=5/144$.
For $\alpha=5/144$ the potential increases if the  node 10  moves to  $S_{16}$.
Now  $ P(S_{17},N\setminus S_{17})=68-272\alpha$ and the new cutoff point is $\alpha=10/289$. Finally, in order to find the upper bound of the resolution parameter, we have to check that for any player there is no incentive to move from her coalition to the empty coalition.

Thus,  for $1/16 \geq \alpha \geq 10/289$  the Nash-stable partition  is
$$
S_{17}=\{1,2,3,4,5,6,7,8,10,11,12,13,14,17, 18,20,22\}\cup \{N\setminus S_{17} \}.
$$
Notice that in this new partition the node 3 belongs to the ``right'' coaltion.

\medskip
Another natural approach to define a symmetric value function is, roughly speaking,
to compare the network under investigation with the configuration random graph model.
The configuration random graph model can be viewed as a null model for a network with
no community structure. Namely, the following value function can be considered:
\begin{equation}
\label{v2}
v_{ij}=\beta_{ij}\left(A_{ij}-\gamma\frac{d_i d_j}{2m}\right),
\end{equation}
where $A_{ij}$ is a number of links between nodes $i$ and $j$,
$d_i$ and $d_j$ are the degrees of the nodes $i$ and $j$, respectively,
$m=\frac{1}{2} \sum_{l\in N} d_l$ is the total number of links in the network,
and $\beta_{ij}=\beta_{ji}$ and $\gamma$ are some parameters.

Note that if $\beta_{ij}=\beta, \forall i,j \in N$ and $\gamma=1$, the potential (\ref{potential2})
coincides with the network modularity \cite{girvan,newman1}. If $\beta_{ij}=\beta, \forall i,j \in N$
and $\gamma \neq 1$, we obtain the generalized modularity presented first in \cite{RB06}.
The introduction of the non-homogeneous weights was proposed in \cite{Wetal10} with the following
particularly interesting choice:
$$
\beta_{ij} = \frac{2m}{d_id_j}.
$$
The introduction of the resolution parameter $\gamma$ allows to obtain clustering with
varying granularity and in particular this helps to overcome the resolution limit \cite{FB07}.

Thus, we have now a game-theoretic interpretation of the modularity function.
Namely, the coalition partition $\Pi=\{S_1,\ldots,S_K\}$ which maximises the modularity
\begin{equation}
P(\Pi)=\sum\limits_{k=1}^K{\sum\limits_{i,j\in S_k, i\neq j}\left(A_{ij}-\frac{d_i d_j}{2m}\right)}
\end{equation}
gives the Nash-stable partition of the network in the Hedonic game with the value function defined by
(\ref{v2}), where $\gamma=1$ and $\beta_{ij}=\beta$.

\medskip

\noindent {\bf Example 1 (ctnd).}
For the network example presented in Fig. 1 we calculate $P(N)=3/2,  P(\{B,C\}\cup \{A,D\}\cup \{E,F\})=
P(\{A,B,C,D\}\cup \{E,F\})=7/2$  and $P(\{A,B,C\}\cup \{D,E,F\})=5$.
Thus, according to the value function (\ref{v2}) with $\gamma=1$ and $\beta_{ij}=\beta$ (modularity
value function), $\Pi=\{\{A,B,C\},\{D,E,F\}\}$ is the unique Nash-stable coalition.
\bigskip

\medskip

\noindent {\bf Example 3 (ctnd).}
Numerical calculations show that the partition $S_{17}\cup \{N\setminus S_{17}\}$ gives the maximum of
potential function (10). It means that this partition is Nash stable.
\bigskip

\section{Conclusion and future research}

We have presented two cooperative game theory based approaches for network partitioning.
The first approach is based on the Myerson value for graph constrained cooperative game,
whereas the second approach is based on hedonic games which explain coalition formation.
We find the second approach especially interesting as it gives a very natural way to
tune the clustering resolution and generalizes the modularity based approaches. Our
near term research plans are to test our methods on more social networks and to develop
efficient computational Monte Carlo type methods.

\section*{Acknowledgements}
This research is supported by Russian
Humanitarian Science Foundation (project 15-02-00352),
Russian Fund for Basic Research (projects 16-51-55006 and 17-11-01079),
EU Project Congas FP7-ICT-2011-8-317672 and Campus France. This is an author edited copy
of the paper published in Proceedings of CSoNet/COCOON 2017.


\bigskip

\begin{thebibliography}{99}


\bibitem{Aetal08}
Avrachenkov, K., Dobrynin, V., Nemirovsky, D., Pham, S.K., and Smirnova, E.:
Pagerank based clustering of hypertext document collections.
In Proceedings of ACM SIGIR 2008, pp.873-874, (2008).

\bibitem{ACN14}
Avrachenkov, K., El Chamie, M., and Neglia, G.:
Graph clustering based on mixing time of random walks.
In Proceedings of IEEE ICC 2014, pp.4089-4094, (2014).

%
%

\bibitem{BWD96}
Blatt, M., Wiseman, S., and and Domany, E.:
Clustering data through an analogy to the Potts model.
In Proceedings of NIPS 1996, pp.416-422 (1996).

\bibitem{Betal08}
Blondel, V. D., Guillaume, J. L., Lambiotte, R., and Lefebvre, E.:
Fast unfolding of communities in large networks.
Journal of statistical mechanics: theory and experiment. v.10, P10008.

\bibitem{Bogomolnaia} Bogomolnaia, A., Jackson, M.O.:
The stability of hedonic coalition structures. Games and Economic Behavior, v.38(2), 201-230 (2002).

%
%
%

\bibitem{D00}
Dongen, S.:
Performance criteria for graph clustering and Markov cluster experiments,
CWI Technical Report (2000).


\bibitem{F10}
Fortunato, S.:
Community detection in graphs.
Physics reports. v.486(3), pp.75-174, (2010).

\bibitem{FB07}
Fortunato, S., and Barthelemy, M.: Resolution limit in community detection.
Proceedings of the National Academy of Sciences, v.104(1), pp.36-41 (2007).

\bibitem{girvan}
Girvan, M., Newman, M.E.J.:
Community structure in social and biological networks. Proc. of National Acad. of Sci. USA, v.99(12),
pp.7821-7826 (2002)

\bibitem{jac2005}
Jackson, M.O.: Allocation rules for network games. Games and Econ. Behav., v.51(1), pp.128-154 (2005)


\bibitem{jac}
Jackson, M.O.: Social and economic networks. Princeton University Press (2008)

\bibitem{Letal09}
Leskovec, J., Lang, K.J., Dasgupta, A., and Mahoney, M.W.:
Community structure in large networks: Natural cluster sizes and the absence of large well-defined clusters.
Internet Mathematics. v.6(1), pp.29-123 (2009).

\bibitem{maz_gt}
Mazalov, V.: Mathematical Game Theory and Applications. Wiley (2014)

\bibitem{fundam}
Mazalov, V.,  Avrachenkov, K., Trukhina,l., and  Tsynguev, B.:
Game-theoretic centrality measures for weighted graphs.
Fundamenta Informaticae. v.145(3), pp.341-358 (2016).

\bibitem{maz_tr}
Mazalov, V.V., Trukhina, L.I.: Generating functions and the Myerson vector in communication networks. Disc. Math. and Appl.
v.24(5), pp.295-303 (2014)

\bibitem{MS01}
Meila, M. and Shi, J.:
A Random Walks View of Spectral Segmentation.
In Proceedings of AISTATS 2001.

\bibitem{my}
Myerson, R.B.: Graphs and cooperation in games. Math. Oper. Res., v.2, pp.225-229 (1977)

\bibitem{newman}
Newman, M.E.J.: A measure of betweenness centrality based on random walks. Proc. of the National Academy of Sciences of the USA,
v.27, pp.39-54 (2005)

\bibitem{newman1}
Newman, M.E.J.: Modularity and community structure in networks. Social networks, v.103,
no.23, pp. 8577-8582 (2006)

\bibitem{PL06}
Pons, P., and Latapy, M.:
Computing communities in large networks using random walks.
Journal of Graph Algorithms and Applications. v.10(2), pp.191-218 (2006).

\bibitem{RAK07}
Raghavan, U. N., Albert, R., and Kumara, S.:
Near linear time algorithm to detect community structures in large-scale networks.
Physical review E. 76(3), 036106, (2007).

\bibitem{RB06}
Reichardt, J., and Bornholdt, S.:
Statistical mechanics of community detection. Physical Review E, 74(1), 016110 (2006).

%
%

\bibitem{L07}
von Luxburg, U.:
A tutorial on spectral clustering. Statistics and Computing, v.17(4), pp.395-416 (2007).

\bibitem{Wetal10}
Waltman, L., van Eck, N. J., and Noyons, E.C: A unified approach to mapping and clustering of bibliometric networks.
Journal of Informetrics, v.4(4), pp.629-635 (2010).

\bibitem{Z77}
Zachary, W.W.:
An information flow model for conflict and fission in small groups.
Journal of anthropological research, v.33(4), pp.452-473 (1977).

\end{thebibliography}
\end{document}